\documentclass[a4paper,10pt]{article}
\usepackage[utf8]{inputenc}
\usepackage{float}
\usepackage{graphicx}
\usepackage[dvipsnames]{xcolor}
\usepackage{hyperref}
\usepackage{amsmath}
\usepackage{subcaption}
\usepackage[export]{adjustbox}
\usepackage{color}
\usepackage[linesnumbered,ruled]{algorithm2e}
\usepackage{longtable}
\usepackage{authblk}
\usepackage{fullpage}
\usepackage{setspace}
\usepackage{sidecap}
\usepackage{pbox}
\usepackage{cite}
\usepackage[capitalise]{cleveref}
\usepackage{tabularx}
\usepackage{placeins}
\usepackage{array, booktabs, makecell,longtable}
\usepackage{nameref}

\usepackage{tabularx}
    \newcolumntype{L}{>{\raggedright\arraybackslash}X}

\DeclareUnicodeCharacter{0394}{\ensuremath{\Delta}}

\makeatletter
\def\BState{\State\hskip-\ALG@thistlm}
\def\arraystretch{1.5}
\makeatother

\title{\bf{Collective Dynamics of Dark Web Marketplaces}}

\author[a,b]{Abeer ElBahrawy}
\author[c]{Laura Alessandretti}
\author[b]{Leonid Rusnac}
\author[b]{Daniel Goldsmith}
\author[d]{Alexander Teytelboym}
\author[a,e,f,*]{Andrea Baronchelli}

\affil[a]{{\small City, University of London, Department of Mathematics, London EC1V 0HB, UK }}
\affil[b]{{\small Chainalysis Inc, NY, USA}}
\affil[c]{{\small Technical University of Denmark, DK-2800 Kgs. Lyngby, Denmark}}
\affil[d]{{\small Department of Economics, University of Oxford, UK}}
\affil[e]{{\small UCL Centre for Blockchain Technologies, University College London, UK}}
\affil[f]{{\small The Alan Turing Institute, British Library, $96$ Euston Road, London NW$12$DB, UK}}
\affil[*]{{\small Corresponding author:  Andrea.Baronchelli.1@city.ac.uk}}

\date{}

\usepackage{makecell}

\begin{document}

\maketitle
\begin{abstract}

Dark web marketplaces are websites that facilitate trade in illicit goods, mainly using Bitcoin. Since dark web marketplaces are illegal, they do not offer any user protection, and police raids and scams regularly cause large losses to marketplaces participants. However, the uncertainty has not prevented the proliferation of dark web marketplaces. Here, we investigate how the dark web marketplace ecosystem re-organises itself following marketplace closures. We analyse $24$ separate episodes of unexpected marketplace closures by inspecting $133$ million Bitcoin transactions among $38$ million users. We focus on users who move their trading activity to a different marketplace after a closure. We find that most migrating users continue trading activity on a single coexisting marketplace, typically the one with the highest trading volume. User migration is swift and trading volumes of migrating users recover quickly. Thus, although individual marketplaces might appear fragile, coordinated user migration guarantees overall systemic resilience. 

\end{abstract}

\section*{Introduction}

Dark web marketplaces (or ``dark markets'') are commercial websites specialised in trading illicit goods. They are accessible via darknets (e.g., Tor) and vary in specialization, technology, and primary supported language. \emph{Silk Road}, the first modern dark marketplace launched in 2011, limited its sales to drugs while other dark marketplaces allow the trading of weapons, fake IDs and stolen credit cards~\cite{gwern,barcclosed}. Most marketplaces simply facilitate transactions between buyers and sellers of illicit goods, however some marketplaces act as sellers and sell directly to buyers. Bitcoin is the universally accepted currency (occasionally together with other cryptocurrencies) on every dark marketplace.  

Operating outside of law, dark marketplaces do not offer any protection to customers or vendors. This has led to a proliferation of scam sales and marketplace hacks. Furthermore, marketplaces may be suddenly closed either by the authorities or by its administrators, causing significant losses to users. For example, \emph{Silk Road} was shut down in 2013 by the FBI~\cite{fbirepoty} and in the same year \emph{Sheep Marketplace} marketplace was closed by its administrator, who vanished with $100$ million US dollars stolen from its users \cite{sheepclosure}. Following these events, dark marketplaces have adopted better technologies to mitigate losses caused by closures and reassure customers \cite{zantout2011i2p,buterin2014bitcoin,wehinger2011dark}. However, this has not prevented further marketplace closures, both due to police seizures and scams.

Counterintuitively, such uncertainty has not prevented a steady growth of both users and revenue of dark marketplaces. As of today, there are at least $38$ identified active dark marketplaces \cite{christin2013traveling}. Although it is difficult to identify relevant transactions from the Bitcoin blockchain and to quantify marketplace volume \cite{christin2013traveling,aldridge2014not,soska2015measuring,dolliver2015evaluating}, the European authorities have estimated that between $2011$ and $2015$ dark marketplace drug sales were $44$ million US dollars per year. A subsequent study estimated that, in early 2016, dark marketplaces drug sales have grown to between $170$ million and $300$ million US dollars per year. \cite{united2018world}. Recently,  \emph{Berlusconi}, known mostly for selling stolen IDs, was seised by the Italian police who estimated their annual transactions to $2$ million euros~\cite{barcclosed}.

Several papers have attempted to study dark marketplaces. However, the difficulty of identifying relevant transactions \cite{christin2013traveling,aldridge2014not,soska2015measuring,dolliver2015evaluating} has forced researchers to rely mostly on user surveys \cite{barratt2014use,van2013silk} and data scraped from dark marketplaces websites \cite{soska2015measuring,decary2017police} (even though dark marketplaces administrators actively fight web scraping which is perceived as a threat). Police shutdowns have been shown to correlate with a sudden increase in drug listings in co-existing marketplaces \cite{van2014closure,buxton2015rise}. The most comprehensive study on closures among $12$ dark marketplaces concluded ``that the effect of law enforcement takedowns is mixed as best'' \cite{soska2015measuring}. Another recent analysis of a large $2014$ police operation identified an impact of closures on the supply and demand of drugs (but not the prices) \cite{decary2017police}. Recent research on attributing anonymised Bitcoin addresses to named entities \cite{tasca2018evolution,ron2013quantitative,meiklejohn2013fistful,
harrigan2016unreasonable} has not yet been applied to the investigation of the dynamics of dark marketplaces.

In this paper, we investigate the dynamics of $24$ marketplace closures by looking at $31$ marketplaces in the period between June $2011$ to July $2019$. We do so by analysing a novel dataset of Bitcoin transactions involving dark marketplaces assembled on the basis of the most recent identification methods \cite{huang2018tracking,harlev2018breaking,goldsmith2019analyzing}. We are therefore able to quantify the overall activity of the major dark marketplaces, in terms of number of users and total volume traded. We show that the closure of a dark marketplace, due to a police raid or an exit scam, has only a temporary effect on trading volumes, suggesting that dark marketplace ecosystem is resilient. We provide the first systematic investigation of dark marketplace users migration following an unexpected closure, and show that closures mainly affect low-active users, with highly-active users migrating quickly to a new marketplace. Finally, we reveal a striking pattern of post-closure coordination: 66\% of migrating users choose to move their activity to the same coexisting marketplace. Moreover, the marketplace that receives the largest number from migrating users tends to have the largest volume and the most users in common with the closed marketplace.

\section*{Methods}
Dark marketplaces operate similar to other online marketplaces, such as eBay, Gumtree or Craigslist, on which vendors advertise their products and price and customers request the shipment through the website and vendors are typically responsible for the delivery. Typically, transactions flow from buyers to the dark marketplace that then sends the money to sellers after buyers confirmation of receiving the goods. Consumers may leave reviews that contribute to vendors' reputation~\cite{wehinger2011dark}. Dark marketplaces are also supported by search engines and news websites such as Grams, DeepDotWeb and darknetlive which aggregate information on all active dark marketplaces~\cite{darknetlive}. After multiple scam closures, nowadays dark marketplaces rely often on escrow systems. The dark marketplace does not keep buyers' bitcoins in local addresses but instead sends it to an escrow service. Escrow services can be independent from the dark marketplace or integrated to the dark marketplace, either way users can withdraw their money (refund it) if the shipment was not delivered. After the buyer's confirmation of receipt, the escrow service transfers the money to the seller. 

Our analysis relies on a novel dataset of dark marketplace transactions on the Bitcoin blockchain.
The ledger of Bitcoin transactions (the blockchain) is publicly available and can be retrieved through Bitcoin core \cite{bitcoincore} or a third-party API such as Blockchain.com \cite{blockchiancom}. It consists of the entire list of transaction records, including time, transferred amount, origin and destination addresses. Addresses are identifiers of $26-35$ alphanumeric characters that can be generated at no cost by any Bitcoin user. Therefore, a single Bitcoin wallet can be associated to multiple addresses. Even though it is called wallet, Bitcoin wallet is better described as a key chain (similar to mac keychain) where users have multiple keys allow them to access multiple addresses where the Bitcoins are stored.  In fact, to ensure privacy and security, most Bitcoin software and websites help users generate a new address for each transaction. In order to be useful, therefore, blockchain data has to be pre-processed to map groups of addresses to individual users.
  
We used data pre-processed by Chainalysis Inc. following the approach detailed in~\cite{huang2018tracking,harlev2018breaking,
goldsmith2019analyzing}. The pre-processing relies on state-of-the-art heuristics \cite{meiklejohn2013fistful,tasca2018evolution,ron2013quantitative,harrigan2016unreasonable,androulaki2013evaluating}, including co-spending clustering, intelligence-based clustering, behavioural clustering, and entity identification through direct interaction~\cite{harlev2018breaking}. These techniques rely on the observation of patterns in the Bitcoin protocol transactions and user behaviour. First, addresses were grouped based on a set of conditions, following some of the heuristics mentioned above and discussed in Supplementary information S$1$. Addresses meeting all conditions were included as part of a single cluster. Note that this step is unsupervised and, there is no ground truth regarding the mapping between addresses and entities \cite{meiklejohn2013fistful}. Then, clusters were identified as specific dark markets, using transaction data collected by Chainalysis (the technique employed for the identification is similar to the one described in~\cite{meiklejohn2013fistful}). Identification of addresses by Chainalysis Inc. related to illicit activities has been relied upon in many law enforcement investigations~\cite{childcase,chung2019cracking}. Given the potential uses of identified Bitcoin data, rigorous investigation and avoidance of false positives is crucial. If an address does not meet all the conditions required by the clustering and identification heuristic, it will be tagged as ``unnamed''. This means that some addresses belonging to a dark marketplace administrator or dark marketplace users, are not included in our dataset (see more information on our dataset in Supplementary Information Section~S$1$).

\begin{figure}
\centering
\includegraphics[width=5in]{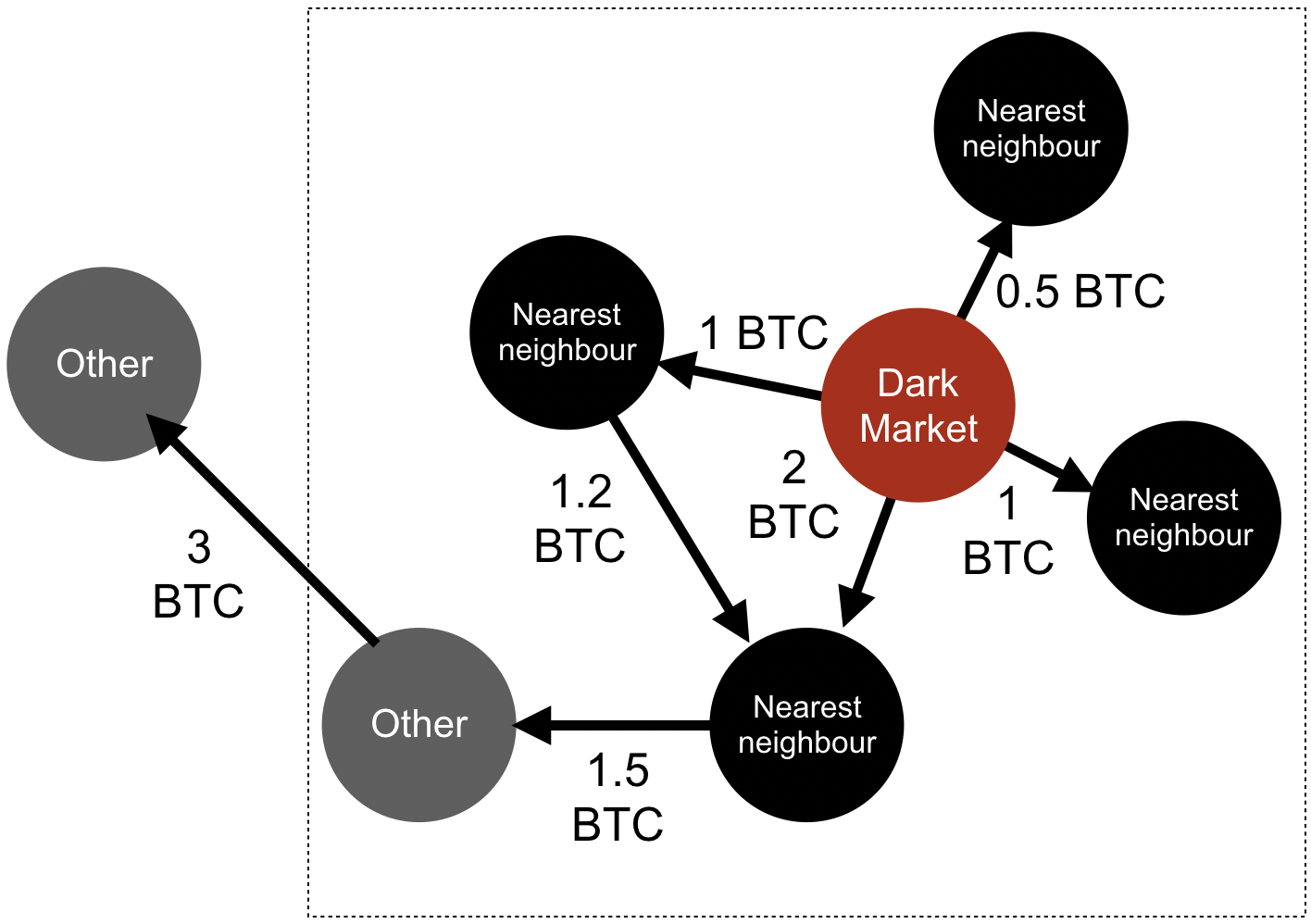} 
\caption{\textbf{Dark market ego-network.} Our dataset includes transaction between addresses belonging to a dark marketplace (in red) and its nearest neighbours (in black), as well as the transactions between nearest neighbours and ``other'' Bitcoin addresses (in grey). Arrows correspond to transactions, and their value in Bitcoin (BTC) is reported. Any transaction between two ``other'' nodes is excluded from our dataset. In this schematic representation, the dotted square includes transactions present in our dataset.}
\label{fig:diff_data_samp}
\end{figure}

We considered the entire transaction data of $31$ dark marketplaces (see Supplementary Information Section~S2) between June $18$, $2011$, and July $24$, $2019$. This dataset includes the major marketplaces on the darknet as identified by law enforcement agencies reports~\cite{europolDrugs,fbirepoty} and the World Health Organization~\cite{united2019world}. We also considered transactions of all users who interacted with one of these marketplaces (dark marketplace's ``nearest neighbours'') after their first interaction with a dark marketplace. Thus, each marketplace can be represented as an egocentric network \cite{marsden2002egocentric} of radius $2$, where the marketplace is the central node, its nearest neighbours represent marketplace users, and “other nodes” appear only through their interaction with one of the marketplaces users. A direct edge represents a transaction occurring either between the marketplace and one of its nearest neighbours, or between two nearest neighbours, or between the nearest neighbour and “other node”. We excluded Bitcoin trading exchanges from our list of nearest neighbours since we focus on the users' direct interaction with the marketplace. Bitcoin trading exchanges are platforms that allow users to trade Bitcoin for other cryptocurrencies or fiat currencies. Fig.~\ref{fig:diff_data_samp} shows a schematic representation of our dataset, where transactions within the square are the ones included in the dataset. After removing transactions to/from cryptocurrency exchanges, the dataset contains $\sim 133$ million transactions among over $38$ million distinct users. The total number of addresses which directly interacted with dark marketplaces is $\sim 8.3$ million. The volume of transactions sent and received by dark marketplaces addresses amount to $\sim 4.2$ billion US dollars. 

\begin{figure}
\centering
\includegraphics[width=5in]{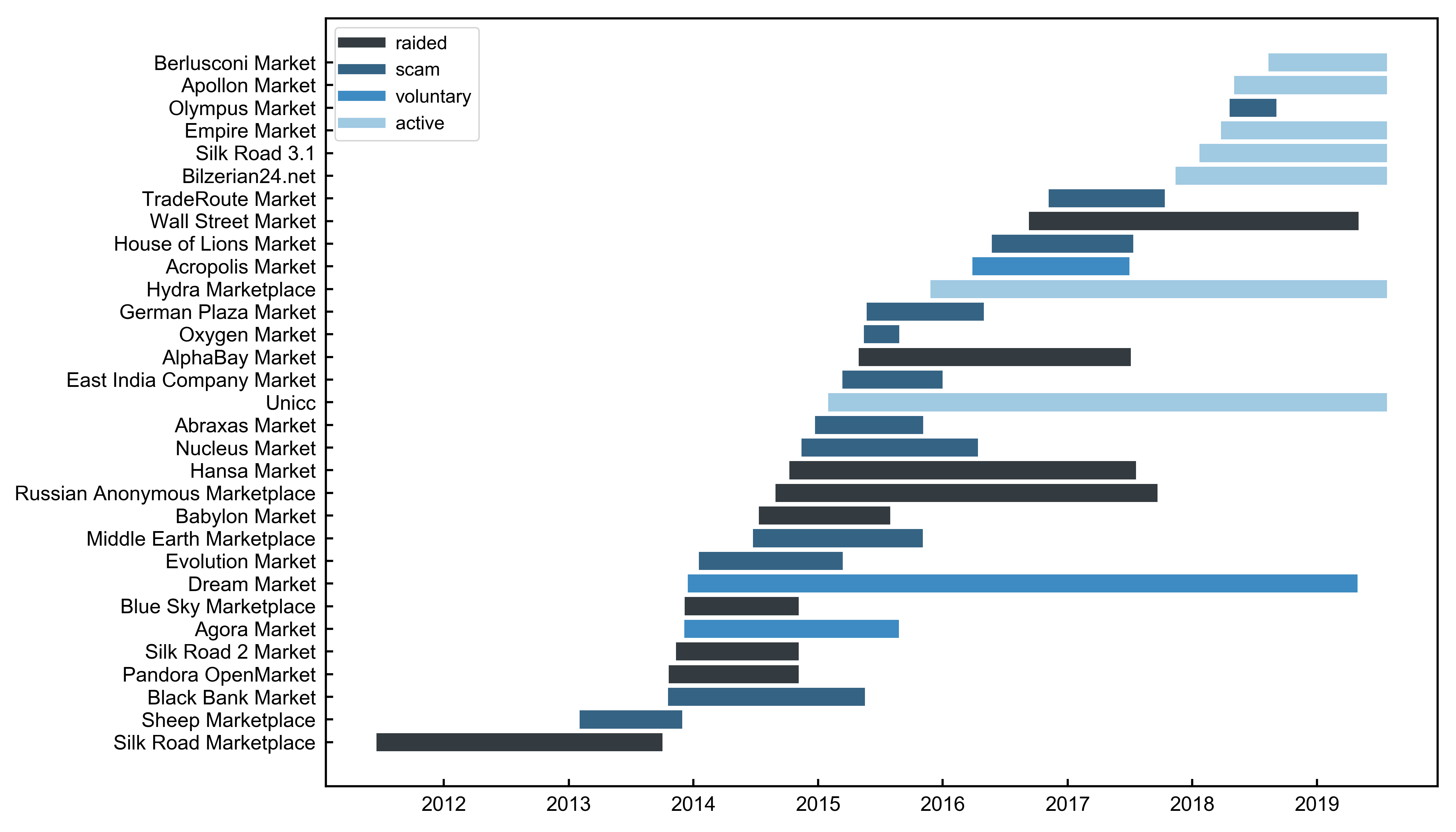}
\caption{\textbf{Dark marketplaces lifetime.} Each bar corresponds to a different dark marketplace (see y-axis labels). Bars are coloured according to the reason behind closure: raided by the police (black), exit scam (dark blue), voluntary closure (blue). Light blue bars correspond to marketplaces that are still active in November $2019$.}
\label{fig:dm_timeline}
\end{figure}

In order to gain information on the analysed marketplaces, we collected additional data from the Gwern archive on dark marketplace closures \cite{gwern}. To compile comprehensive information, we also used law enforcement documents on closures as well as a number of online forums~\cite{europolDrugs,united2019world,zamani2019differences} dedicated to discussing dark marketplaces (see Supplementary Information Section~S2). Out of the selected marketplaces, $12$ were subject to an exit scam, $9$ were raided, $3$ were voluntarily closed by their administrators, and $7$ are still active. $29$ marketplaces operate in English and $2$ operate in Russian. Out of the $31$ marketplaces, $3$ are marketplaces dedicated to fake and stolen IDs and credit cards. The primary currency on these marketplace is Bitcoin. In Fig.~\ref{fig:dm_timeline}, we present the lifetime of the selected marketplaces and the reason behind their closure.

\section*{Results}

The dataset contains $133,308,118$ transactions among $38,886,758$ users. The total number of distinct users which directly interacted with a dark marketplace is $8,377,478$. The volume of transactions sent and received by dark marketplace addresses amount to $4.210$ billion US dollars, while the one received by dark marketplaces address is $1.99$ billion US dollar. Note that the conversion between Bitcoins and US dollars is done considering the price of Bitcoin at the time of the transaction. Table~S2 in Supplementary Information Section S2 reports characteristics of the $31$ marketplaces considered, including overall number of users and transaction volume. The most active marketplace in terms of number of users and traded volume is \emph{AlphaBay}, followed by \emph{Hydra}.

\subsection*{Marketplace resilience}

\begin{figure}
\centering
\includegraphics[width=5in]{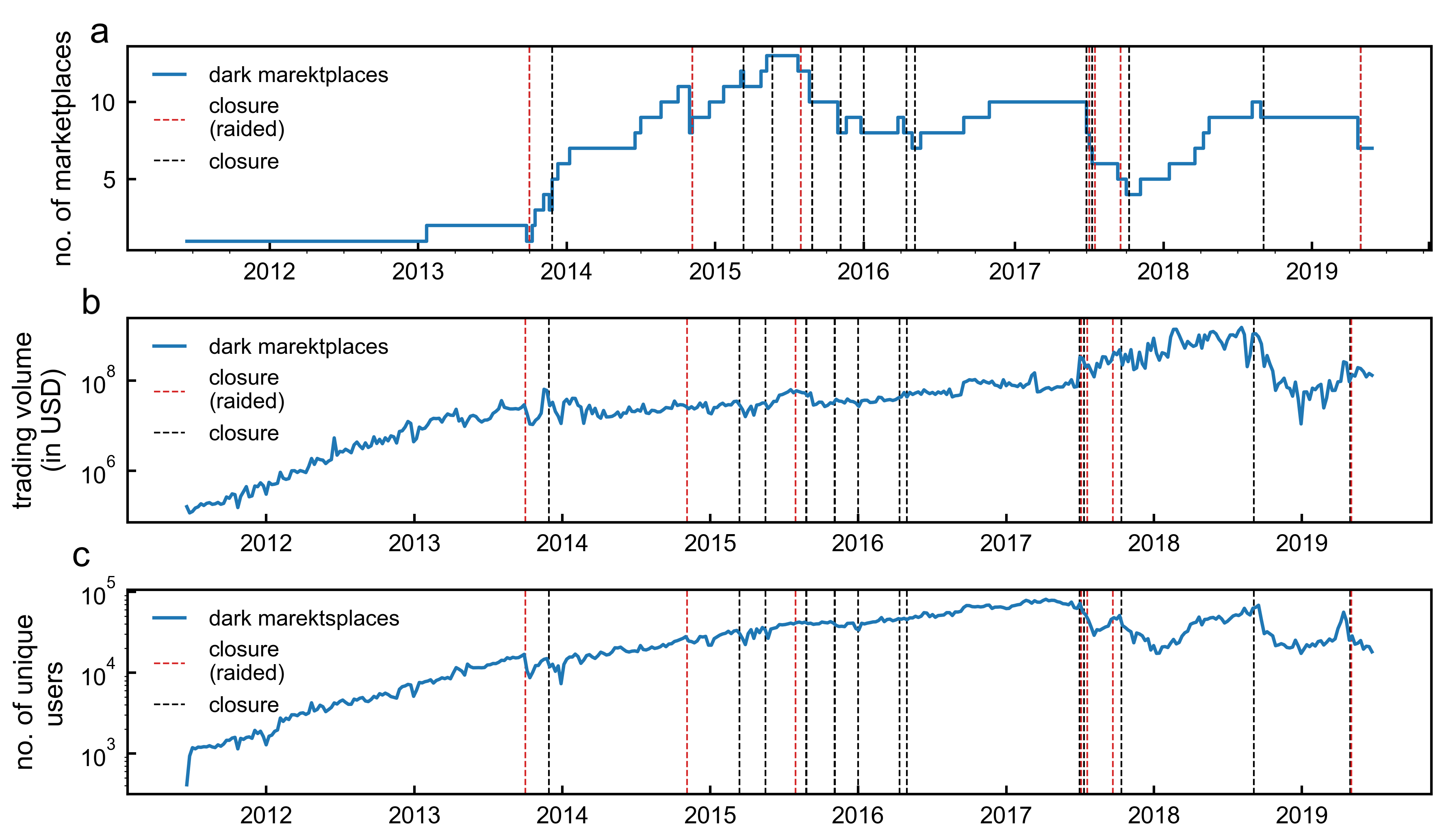} 
\caption{\textbf{Dark marketplaces resilience.} (\textbf{a}) The total number of active dark marketplaces across time. (\textbf{b}) The total volume (in US dollars, USD) exchanged by dark marketplaces addresses. (\textbf{c}) The number of unique users interacting with dark marketplaces. Dashed lines represent marketplace closure due to law enforcement raid (in red), or any other reason (in black). Values are calculated using a time window of one week.}
\label{fig:total_usd}
\end{figure}

The capacity of the dark marketplace ecosystem to recover following the marketplace closure can be studied by quantifying the evolution of the total volume traded by dark marketplaces over time. 
Despite recurrent closures, we find that the number of marketplaces has been relatively stable since $2014$ as new marketplaces frequently open (see Fig.~\ref{fig:total_usd}a). In addition, despite closures, the total weekly volume sent/received by dark marketplace addresses has grown from $2014$ until the end of $2019$ (see Fig.~\ref{fig:total_usd}b). In fact, Moving Average Convergence Divergence (MACD) analysis~\cite{appel1979moving} reveals that, following each dark marketplace closure, the overall dark marketplace volume drops, but it recovers quickly thereafter, within $9$ days on average (median: $3$ days, see also Figure~S$6$ in Supplementary Information). Starting from the end of $2018$, however, we observe a decrease in the total volume traded (See Fig.~\ref{fig:diff_data_samp}).

\subsection*{User migration}

The observation that trading volumes recover quickly after unexpected marketplace closure suggests that users may move to other marketplaces \cite{decary2017police,duxbury2018building}. We refer to this phenomenon as \emph{migration}.

In fact, migration was observed \cite{MartinBlog} after the closure of the \emph{AlphaBay} marketplace when other marketplaces, namely \emph{Hansa} and \emph{Dream Market}, experienced an abnormal spike in activity. In this section, we provide the first systematic investigation of dark marketplace users migration, by studying the effects of a series of closures. 

\begin{figure}
\centering
\includegraphics[width=5in]{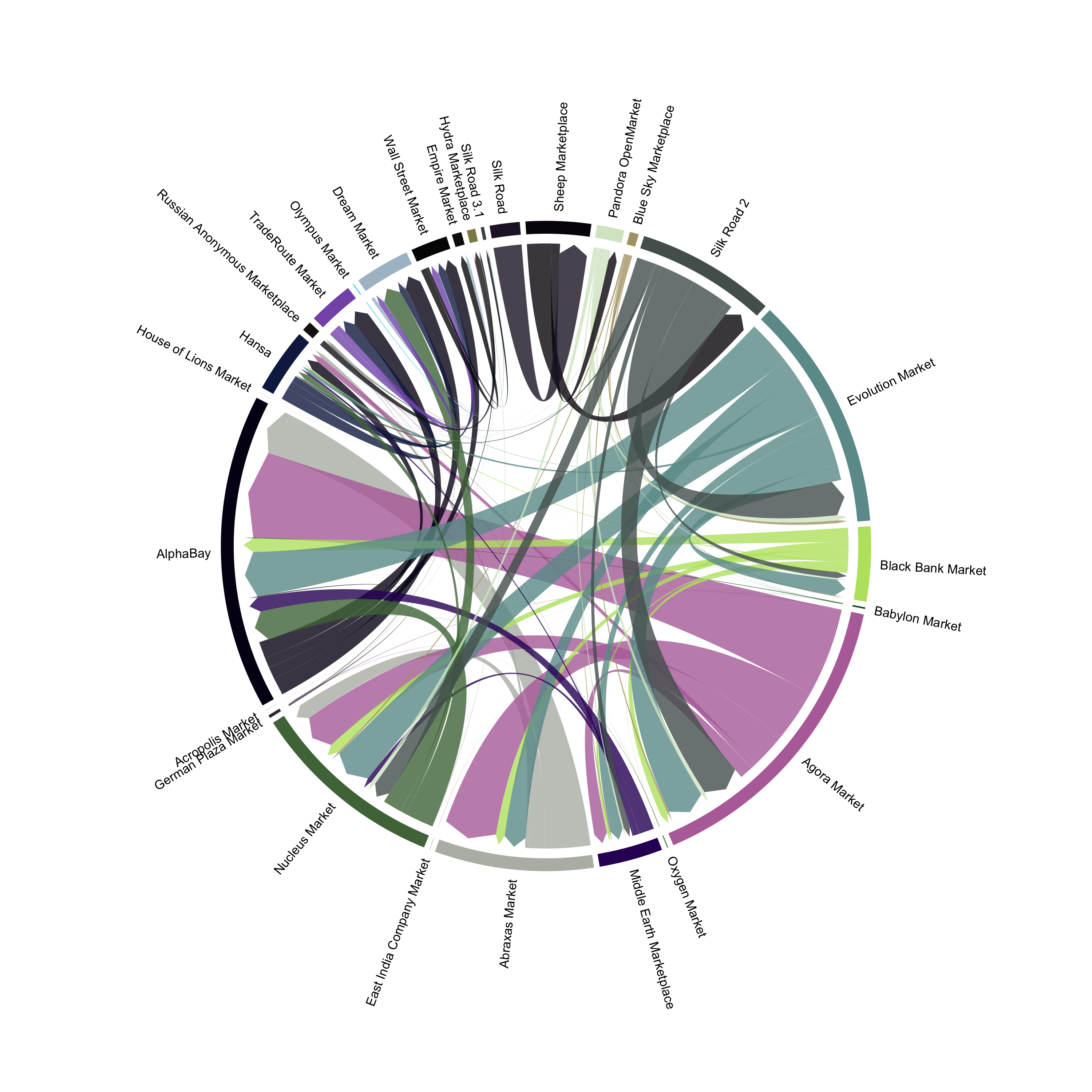} 
\caption{\textbf{Migration of users following a dark marketplace closure.} Flows of users migrating to another coexisting marketplace following a closure. The arrowhead points to the direction of migration, and the width of the arrow represent the number of users. Marketplaces are ordered clockwise according to the closing date in ascending order starting from Silk Road Marketplace.}
\label{fig:chord_mig}
\end{figure}

We identify migrant users in the following way. For each dark marketplace $m$ that was shut down, we identify users who started trading on another coexisting marketplace $m'$ \textit{following} the closure of $m$. If a user was trading on both marketplaces $m$ and $m'$ before the closure of marketplace $m$, the user is not labelled as a migrant to marketplace $m'$.  Fig.~\ref{fig:chord_mig} shows the flows of migrant users between marketplaces.

But what fraction of users do migrate after a closure? To answer this question we need to consider that $\sim 38\%$ of all users in our dataset made only one transaction (sent or a received once) - a finding consistent with the evidence that most of the minted Bitcoin were accumulated in addresses which never sent, at least until 2013 \cite{ron2013quantitative}). We then need to estimate the expected number of users who would have kept transacting with the market if no closure had occurred. To do so, we focus on the ``returning users'' over time, i.e., the fraction of all users active (sent or received) in a given week that are active also in the following day. We denote as $R_t$ the intersection between the set of active users on a given day $t$, and the set of users that were active at least once in the preceding week. Thus, $R_t$ is the set of "returning users". In order to compare the number of returning users across closures, we normalize the entire time series by the fraction of returning users at the time of closure, $R_{\hat{t}}$ where $\hat{t}$ is the day of closure. Thus, the normalized value of returning users on the day of closure is $1$. Then, we consider the median across marketplace closures. We find that, $5$ days after the closure of a dark marketplace, $85\%$ of the expected number of returning users interact with another marketplace, where the `expected' number is computed by considering the typical number of returning users before closure. This finding indicates that, even though the marketplace closure does affect participation, the vast majority of returning users do migrate to another marketplace following a closure.

\subsection*{Who is migrating?}
\begin{figure}
\centering
\includegraphics[width=5in]{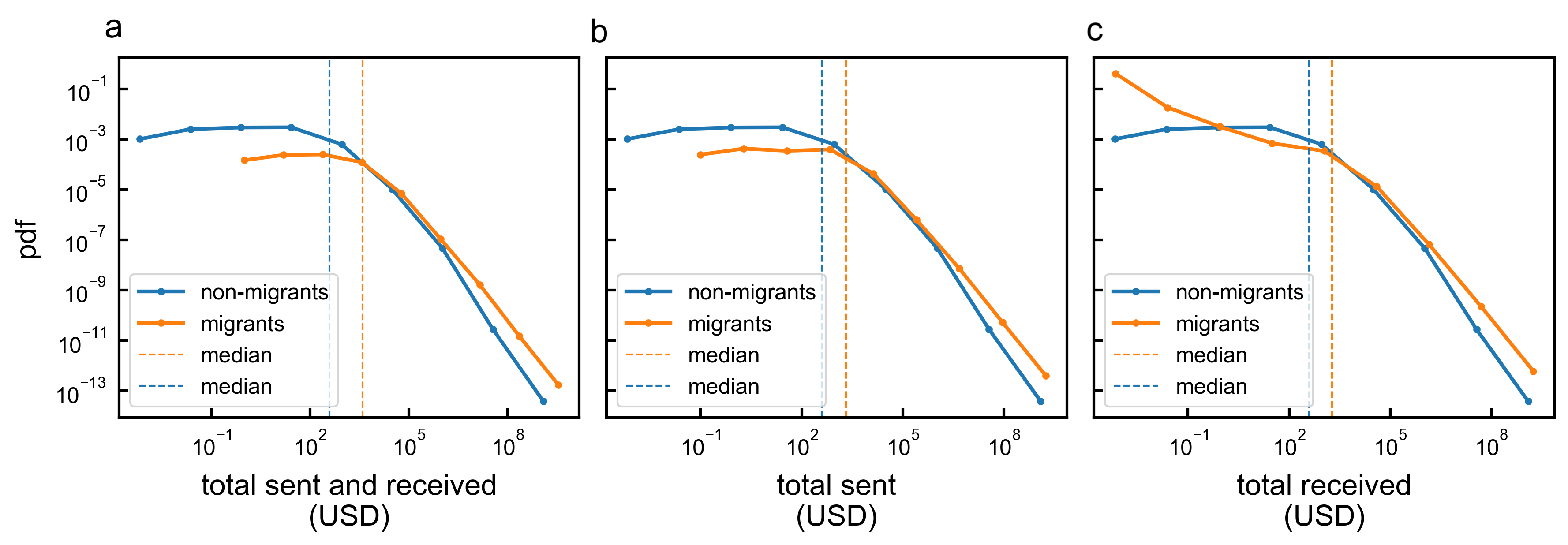} 
\caption{\textbf{Migrant vs. non-migrants activity distribution.} \textbf{(a)} The distribution of the total volume sent and received across all closed dark marketplaces for migrants (orange line) and non migrants (blue line). \textbf{(b)} The distribution of the total volume sent across all closed dark marketplaces by migrants (orange line) and non migrants (blue line). \textbf{(c)} The distribution of the total volume received across all closed dark marketplaces by migrants (orange line) and non migrants (blue line). Dashed lines represent the median value for migrant users (orange line) and non-migrant users (blue line).}
\label{fig:mig_non_dist}
\end{figure}

The observation that some users stop trading following a dark marketplace closure but the total volume traded in dark marketplaces does not decrease could indicate that migrant users are on average more active than others. 
We test this hypothesis by computing the activity of migrant users before and after marketplace closure. We refer to the original dark marketplace that a user was interacting with as its \emph{home marketplace}. 
For all users (migrant and non-migrant), we measure the total volume exchanged with any other user in our dataset including the home marketplace. We find that the median volume exchanged by migrant users is $\sim 10$ times larger than the volume exchanged by non-migrant users (see Fig.~\ref{fig:mig_non_dist}a), with the median volume exchanged summing to $3,882.9$ US dollars across all migrant users and to $387.2$ US dollars for non-migrant users. The means are sensitive to high volume users with $71,6441.9$ US dollars and $17,529.7$ US dollars for migrant and non-migrant users respectively (see Fig.~\ref{fig:mig_non_dist}). In terms of receiving and sending behaviour migrants users are also more active compared to non-migrants (see Figs.~\ref{fig:mig_non_dist}b and c)
Similar conclusions can be drawn by considering the volume exchanged with the home marketplace only,
which has median value of $263$ US dollars for migrant users and for non-migrant users $74.3$ US dollars and a mean value of $2,725.1$ US dollars and $475.9$ US dollars for migrant and non-migrant users respectively (see Supplementary Information Section~S4).

The activity distribution of migrants is significantly different from the non-migrant users' distribution (using Kolmogorov–Smirnov test, $p < 0.01$, see Table~S3 in Supplementary Information Section~S4).

\subsection*{Coordination in the dark}

We now turn to the analysis of how migrant users decide where to migrate.
In our dataset, following every instance of marketplace closure except one, users could migrate to two or more co-existing marketplaces. 

\begin{figure}
\centering
\includegraphics[width=5in]{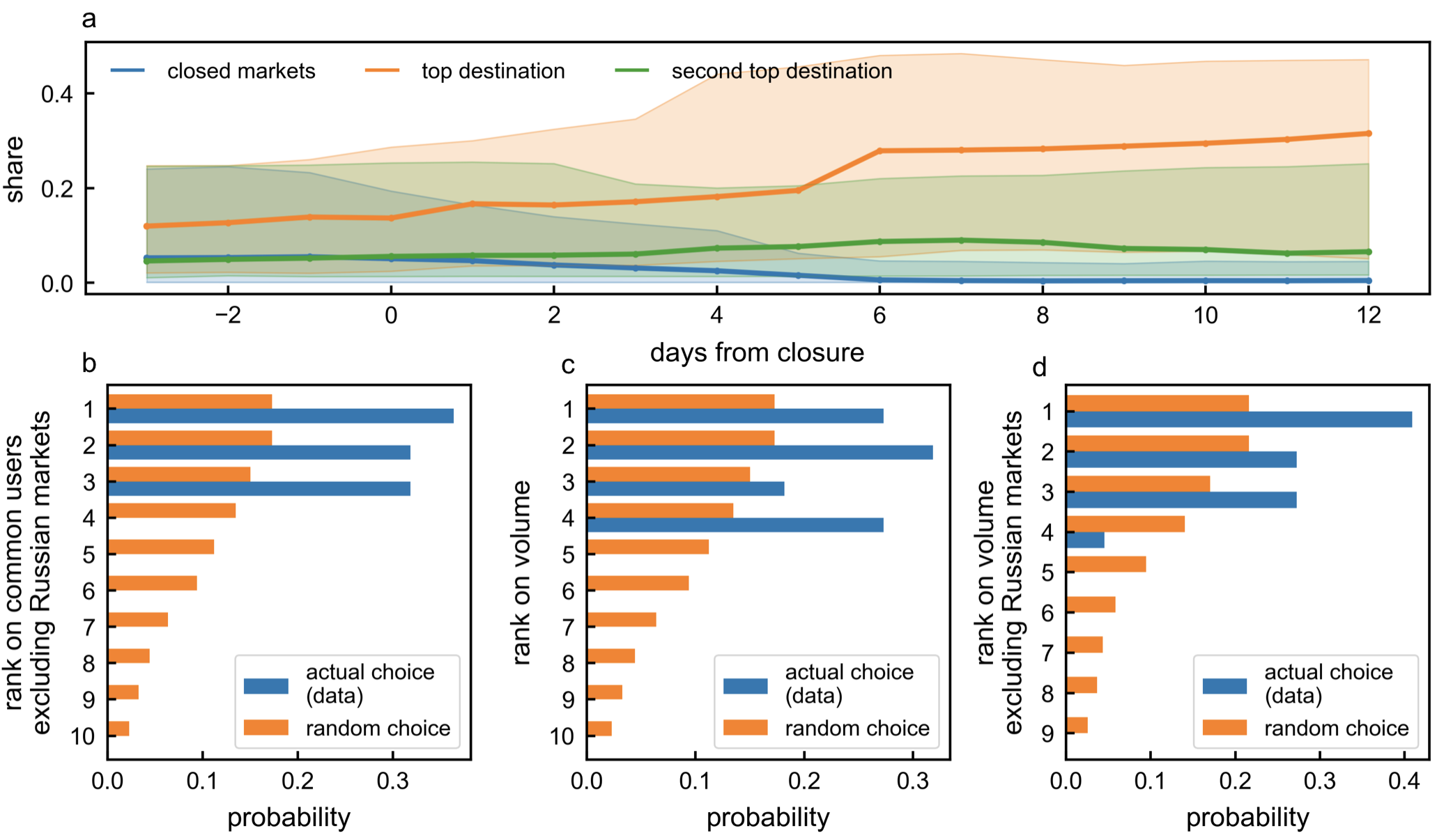} 
\caption{\textbf{Migration decision and impact.} (\textbf{a}) The median share (across closures) of a closed marketplace (blue line), the top destination marketplace for the migrant users (orange line) and the second top destination for migrant users (green line). The shaded area represents the $50\%$ interquartile range. Value are computed using a rolling window of one week. Figures (\textbf{b-d}) show the probability of a marketplace to be chosen for migration (becoming the top destination for migration) given its rank at the time of coexisting marketplace closure in comparison to the random model. Marketplaces are ranked in descending order according to (\textbf{b}) the number of overlapping users they have with the closed marketplaces excluding Russian marketplaces (\textbf{c}) the total trading volume in US dollars and (\textbf{d}) the total trading volume in US dollars excluding Russian marketplaces from the ranking. The random model in the Figures b-d represent a model where users can move to any existent marketplace with equal probability.}  
\label{fig:market_res}
\end{figure}

Fig.~\ref{fig:market_res}\textbf{(a)}, we show the evolution of the trading volume shares of the closed marketplace and the top two destination marketplaces in the days preceding and following a closure. Trading volume share for a give market is the normalised trading volume of a market by the total trading volume of all dark marketplaces. We find that the top two destination marketplaces experience an increase in the trading volume share starting $2$ days after the closure, and saturating about $6$ days after with a share of $27\%$, approximately more than double the share at the time of closure. The second top destination on the other hand its share increases from $5\%$ to $8.7\%$. 

We investigate the characteristics of the top destination marketplace for migrant users, by ranking coexisting marketplaces according to the total trading volume in US dollars at the time of closure and the total number of common users between the shut down and the coexisting marketplace before closure.
We find that, regardless of the reason behind closure, users do not migrate randomly and chose to move to the marketplace with the highest trading volume which, in some cases, is also the marketplace with the highest number of common users.

Focusing on the first week after closure, we find that, on average, one marketplace absorbs $66.1\% \pm 16.1$ of all migrant users. Only $4\%$ of the users migrate to more than one coexisting marketplace simultaneously after the closure. What is this marketplace? Fig.~\ref{fig:market_res}b shows that, in $36.4\%$ of the closures considered, it is the one sharing the larger number of common users with the closed marketplace, while the chances that users select the second and the third rank is $31.8\%$. Users do not choose to migrate to marketplaces with rank lower than the third. 

Fig.~\ref{fig:market_res}c shows that, when marketplaces are ranked according to the volume of their transactions, the second-largest marketplace is preferred in the majority of cases ($31.8\%$). However, a closer look at the data reveals that the Russian marketplace occupies often the top ranks in terms of volume but it tends not to be the preferred migration harbour, probably due  language and geographical barriers. Excluding the Russian marketplace from the ranking, in fact, we find that the largest marketplace by volume is selected $41\%$ of the times (see Fig.~\ref{fig:market_res}d). 

We compare the users' decisions with a null random model, where at each closure users move with equal probability to any of the existent marketplaces. The probability $P_i$ of the $i^\text{th}$-ranked marketplace to be chosen for migration uniformly at random after $m$ closures is equal to $$P_i = \frac{\sum^{m}_{j=1}{1/c_j}}{m},$$ where $c_j$ is the number of coexisting marketplaces at the time of closure $j$. We find that the data different significantly from the uniform random choice model, confirming the presence of coordination between migrating users (see Fig.~\ref{fig:market_res})

\section*{Conclusion}

We analysed a novel dataset of Bitcoin transactions on $31$ large dark marketplaces and investigated how the darknet marketplace ecosystem was affected by the unexpected marketplace closures between $2013$ and $2019$. The marketplaces we considered were heterogeneous in many ways and $24$ of them were closed abruptly due to police raids and scams. We found that the total volume traded on these dark marketplaces dropped only temporarily following closures, revealing a remarkable resilience of the marketplace ecosystem. We identified the origin of this resilience, by focusing on individual users, and unveiled a swift and ubiquitous phenomenon of migration  between recently closed marketplaces and other coexisting ones. We found that migrating users were more active in terms of total transaction volume compared to users who did not migrate. Finally, we found that migrating users tended to migrate predictably to co-existing marketplaces which had the largest overall volume and the most numbers of users in common with the closed marketplace. 

Our findings shed new light on the consequences of sudden closure and/or police raids on dark marketplace, which had been previously raised in the literature and among law enforcement entities~\cite{europolDrugs,united2018world,decary2017police}. Interesting future research directions include the role of marketplace closure on the emergence of new marketplaces, refining the analysis to investigate whether scam closures and police raids may have so-far neglected effects on user migration, delving deeper into the types of user behaviour that can predict migration, and broadening the research to include the effect of online forums on the performance of existing marketplaces as well as on the migration choices after a closure~\cite{zamani2019differences}. More broadly, we anticipate that our findings will help inform future research on the self-organisation of emerging online marketplaces.

\section*{Authors contribution}

A.E., L.A., A.T. and A.B. designed the research; L.R. acquired the data. L.R. and A.E. prepared and cleansed the data, A.E. performed the measurements. A.E., L.A., D.G., A.T. and A.B. analysed the data. A.E., L.A., A.T. and A.B. wrote the manuscript. All authors discussed the results and commented on the manuscript.

\section*{Data and materials availability}

All data needed to evaluate the conclusions in the paper are present in the paper. Additional data related to this paper may be requested from the authors.

\section{Appendix}

\section{Clustering techniques}
\label{app:clust_tech}

\begin{figure}[H]
\centering
\includegraphics[width=5in]{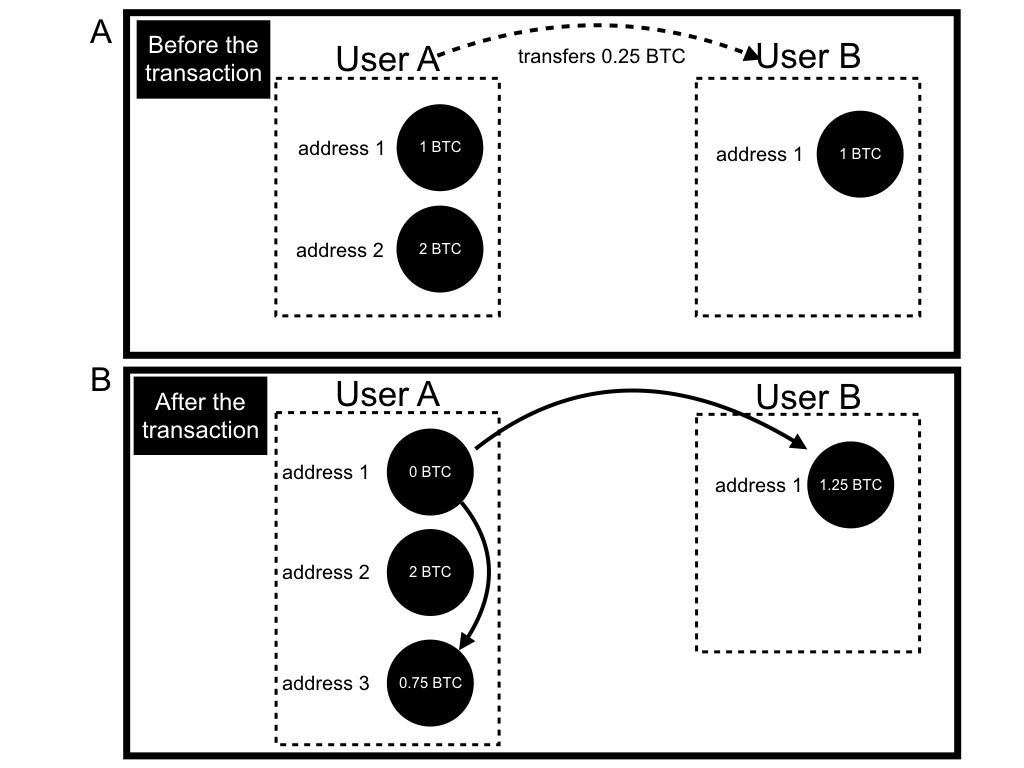} 
\caption{\textbf{How Bitcoin's protocol handles transactions with change.} (\textbf{A}) A transaction between users $A$ and user $B$, where $A$ wants to transfer $0.25$ Bitcoins to $B$. User $A$ has two addresses, one with $1$ Bitcoin and the other with $2$ Bitcoins. User $B$ has one address, containing $1$ Bitcoin. (\textbf{B}) How a transaction is conducted under Bitcoin protocol. User $A$ first address transfers $0.25$ Bitcoin to user $B$ first address. The change of $0.75$ Bitcoin does not stay in User A first address $1$, but appears, instead, as another transaction to a new address. The dotted boundaries in both figures represent a grouping of these addresses, as they belong to one user. A solid arrow represents an executed Bitcoin transaction, while the dotted arrow represents a desired transaction.}
\label{fig:il_change}
\end{figure}

In Bitcoin, multiple addresses can belong to one user; grouping these addresses reduces the complexity of the ledger and Bitcoin anonymity~\cite{ron2013quantitative}. Clustering techniques rely on how Bitcoin's protocol works, users behaviour on the blockchain, Bitcoin's transaction graph structure and finally, machine learning. 
Methods relying on Bitcoin's protocol specifically exploit what is known as change addresses: Bitcoins available in an address have to be spent as a whole. Fig.~\ref{fig:il_change} shows an example of a change address. User A's wallet has two addresses, one contains $1$BTC and another has $2$BTC. User $A$ would like to transfer $0.25$BTC to user $B$, as shown in Fig.~\ref{fig:il_change}A.
After transferring the $0.25$BTC to $B$, the change ($0.75$BTC) will not stay in the same address. Bitcoin protocol will create another address, also assigned to $A$, where the $0.75$BTC change will be stored. By observing this pattern, a heuristic technique proposed in~\cite{androulaki2013evaluating} suggests that these addresses can be grouped, as they belong to one user.

\begin{figure}[H]
\centering
\includegraphics[width=5in]{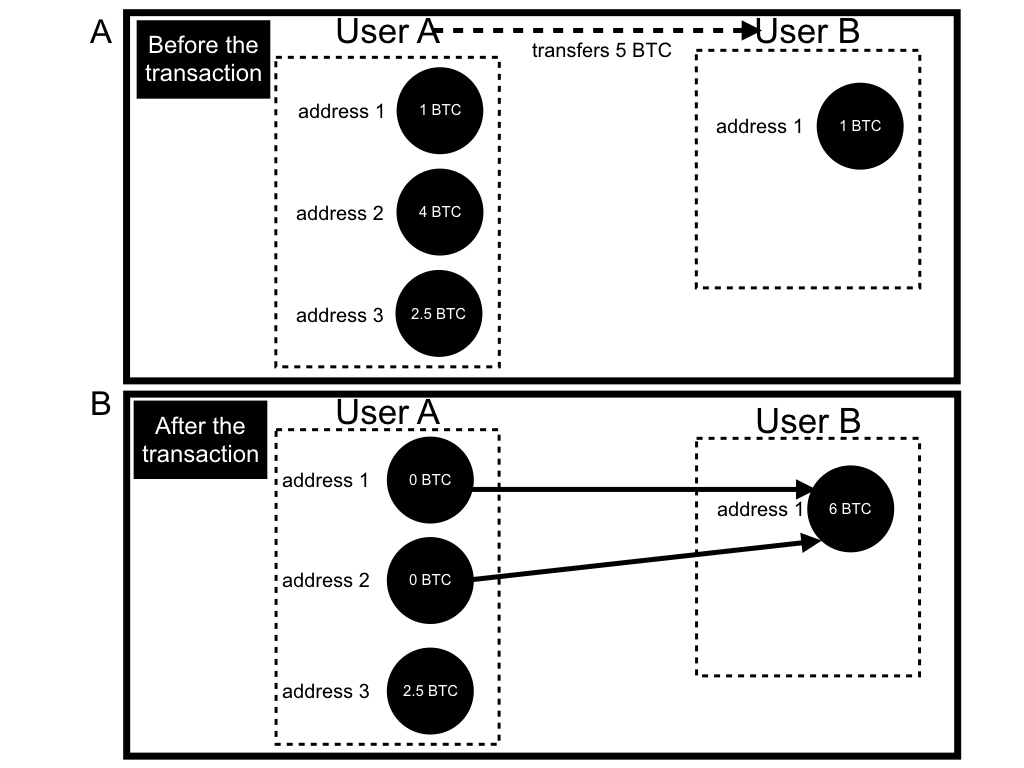} 
\caption{\textbf{Sending from multiple inputs in Bitcoin} (\textbf{A}) A desired transaction between users $A$ and $B$, where $A$ wants to send $5$ Bitcoins to user $B$. User $A$ has $3$ different addresses with $1$, $4$ and $2.5$ Bitcoins respectively. User $B$ has one address containing $1$ Bitcoin. (\textbf{B}) How the transaction will be conducted under the Bitcoin protocol. User $A$ will use two addresses to complete the transaction. Both addresses will send to one address belonging to user $B$. The dotted boundaries in both figures represent a grouping of these addresses as they belong to one user. The solid arrows represent an already executed Bitcoin transaction while the dotted arrow represents a desired transaction.}
\label{fig:il_minputs}
\end{figure}

Since users can have multiple addresses, they can use multiple of these addresses to transfer Bitcoins in a single transaction. For example, Fig.~\ref{fig:il_minputs}A shows a case where user $A$ controls $3$ different addresses. Each address has a different amount of Bitcoins, $1$, $4$ and $2.5$ respectively. User $A$ wants to transfer $5$ Bitcoins to user $B$, and two addresses will be used to complete the transaction as shown in Fig.~\ref{fig:il_minputs}B. This observation allows the grouping of these two addresses as a single user~\cite{androulaki2013evaluating}. 

The work in~\cite{meiklejohn2013fistful} challenged these heuristics, showing the possibility of having false positives and not taking into consideration changes in the protocol. The work suggests instead a manual process, where the behaviour of each entity is investigated. Page rank (network centrality measure~\cite{heidemann2010identifying}) was also used to identify important addresses~\cite{fleder2015bitcoin}; however, the addresses were already grouped using the heuristics introduced by~\cite{androulaki2013evaluating}. Machine learning was also shown to identify addresses which should be grouped as one with $77\%$ accuracy.

Mapping addresses to an actual identity is more challenging. Some entities already publish their public key for donation and payment, such as Wikimedia Foundation~\cite{wikiBTCdon}. The only research that introduced a method for mapping a collection of addresses to a real-world identity is~\cite{meiklejohn2013fistful}, through direct interaction with the address. In this work, researchers directly engaged in $344$ transactions with different services including mining pools, exchanges, dark marketplaces and gambling websites.

The introduction of these heuristics did not only challenge Bitcoin's anonymity but also eased the regulation of Bitcoin. Companies specialising in blockchain analytics started to capitalise on these heuristics and provide tools for exchanges and law enforcement entities to facilitate regulatory efforts. For our analysis of dark marketplaces, our data was provided by  Chainalysis~\cite{chainalysis}, which is a blockchain analytics company. Chainalysis aided several investigations led by different law enforcement entities, including the United States Internal Revenue Service (IRS)~\cite{chung2019cracking}.

\begin{figure}[H]
\centering
\includegraphics[width=5in]{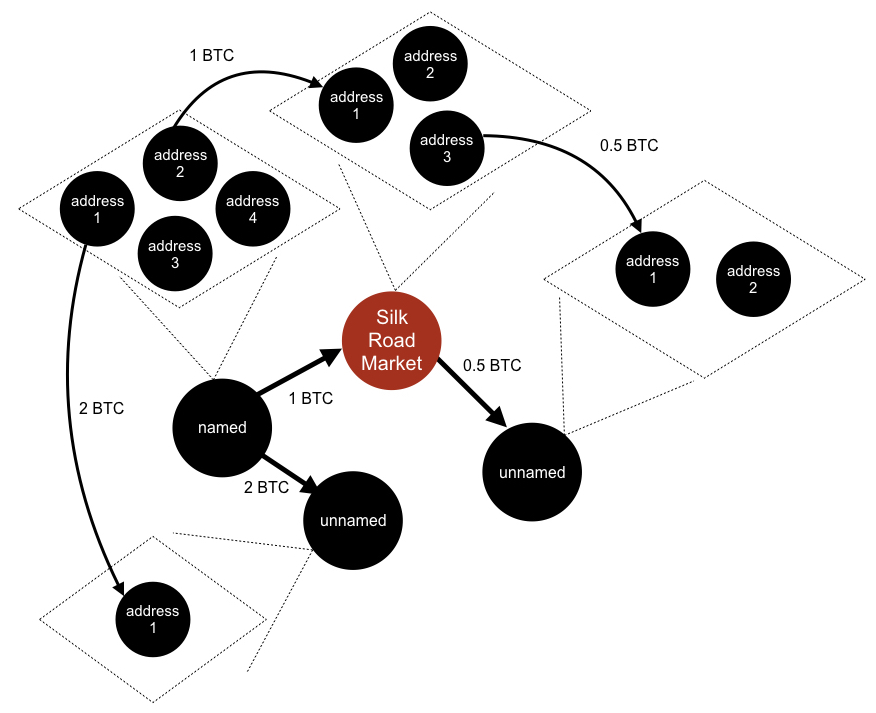} 
\caption{\textbf{A dark marketplace's Bitcoin transaction network.} A schematic representation of our dataset as a complex network. Nodes represent users, and a direct edge between two nodes represents a transaction in the direction of the edge. Nodes can represent different abstractions as shown by the dotted rhombus. Starting from the right side, the unnamed black node represents a cluster of two different addresses which, however, was not attributed to a specific entity. The dark marketplace node(in dark red, Silk Road Market), is a representation of $3$ addresses and attributed by the algorithm to the marketplace. The black named node on the left side of Silk Road Market node is a representation of $4$ addresses and named to belong to a specific entity. Finally, the black unnamed node at the bottom left side of the figure, represents one address.}
\label{fig:diff_nodes_mean}
\end{figure}

Our dataset sampling approach (from the entire Bitcoin transactions) deploys a complex network perspective. Transactions on the blockchain can be modelled as a directed weighted graph where a node represents a user, and a directed edge between two nodes $A$ and $B$ represents a transaction from user $A$ to user $B$. 
Depending on the clustering algorithm, a node can represent one address or multiple addresses. A node can also be labelled as a specific entity or unlabelled (unnamed). Fig.~\ref{fig:diff_nodes_mean} shows a sketch of the network and the different possible meanings of a node. For example in Fig.~\ref{fig:diff_nodes_mean}, the black unnamed node on the right side is a representation of two different addresses clustered together, however, they were not attributed to an entity thus remained unnamed.

\section{Dark marketplaces information}
\label{app:markets_info}

In this section we provide data on each marketplace understudy. Table~\ref{tab:table_darkmarkets} shows general information on the dark marketplaces included in our dataset. 

\begin{table}
\renewcommand\arraystretch{1.5}
\centering
\begin{tabularx}{\columnwidth}{@{}Xllll@{}}
\midrule

{Name} & {Start date} & {End date} & {Closure reason}  &  {Sales} \\
\hline

Abraxas Market    &    $2014-12-13$ &    $2015-11-05$ &        scam    & drugs\\
Acropolis Market    &    $2016-03-27$   & $2017-07-01$     &    voluntary    & mixed\\
Agora Market & $2013-12-03$ & $2015-08-26$ &    voluntary    & mixed\\
AlphaBay Market    & $2014-12-22$ & $2017-07-05$ &    raided    & mixed\\
Apollon Market    &    $2018-05-03$ &   active &    active    & drugs\\
Babylon Market  & $2014-07-11$ & $2015-07-31$ &    raided    & drugs\\
Berlusconi Market        & $2018-08-12$ & active &    active    & mixed\\
Bilzerian24.net    &    $2017-11-13$ &   active &        active    & credits\\
Black Bank Market    &    $2014-02-05$ &    $2015-05-18$ &    scam    & mixed\\
Blue Sky Marketplace    & $2013-12-03$ &    $2014-11-05$ &    raided    & drugs\\
Dream Market    &    $2016-03-19$  & $2019-04-30$ & voluntary &    mixed\\
East India Company Market    &    $2015-04-28$ &    $2016-01-01$ &    scam &    drugs\\
Empire Market    & $2018-02-01$ &    active &    active & mixed\\ 
Evolution Market    &    $2014-01-14$ &    $2015-03-14$ &     scam &    drugs\\
German Plaza Market & $2015-05-22$	& $2016-05-01$ & scam &	mixed\\
Hansa Market    &    $2014-03-09$ &    $2017-07-20$ &    raided    & drugs\\
House of Lions Market    &    $2016-05-23$ &    $2017-07-12$ &    raided    & drugs\\
Hydra Marketplace	& $2015-11-25$ & active & active	 & mixed \\
Middle Earth Marketplace    &    $2014-06-22$ &    $2015-11-04$ & scam & mixed\\
Nucleus Market    &    $2014-10-24$ &    $2016-04-13$ &    scam & mixed\\
Olympus Market        & $2018-04-20$ & $2018-09-04$ & scam & mixed\\
Oxygen Market    &    $2015-04-16$ &    $2015-08-27$ &    scam &    drugs\\
Pandora OpenMarket    &    $2013-10-20$ &    $2014-11-05$ &    raided    & drugs\\
Russian Anonymous Marketplace	&	$2014-08-29$ & $2017-09-21$ & raided & mixed\\
Sheep Marketplace    &    $2013-02-28$ &    $2013-11-29$ &    scam    & drugs\\
Silk Road Marketplace	& $2011-01-31$ & $2013-10-02$ & raided & mixed\\
Silk Road 2 Market    &    $2013-11-06$ &    $2014-11-05$ & raided & mixed\\
Silk Road 3.1	& $2018-01-21$ &	active &	active&	drugs\\
TradeRoute Market    &    $2016-11-06$ &    $2017-10-12$ &    scam    & mixed\\
Unicc	&$2015-01-30$ & active & 	active &credits \\
Wall Street Market	&	$2016-09-09$ &	$2019-05-02$ &	raided	& mixed\\
\bottomrule

\end{tabularx}
\caption{\label{tab:table_darkmarkets} \textbf{Dark marketplaces information.} Information on the $31$ selected dark marketplaces included in our dataset. For each marketplace, the table states the name of the marketplace, the start and end dates of its operation, the closure reason (if applicable) and the type of products sold by the marketplace. ``Drugs'' indicates that the primary products sold on the marketplace are drugs while ``credits'' indicates the marketplace specialises in fake IDs and credit cards and ``mixed'' indicates the marketplace sells both types of products}
\end{table}

Table~\ref{tab:markets_sales} shows the total volume received and sent by the different marketplaces, as well as the number of their users.

\begin{table}
\centering
\renewcommand\arraystretch{1.5}
\begin{tabularx}{\columnwidth}{@{}lrrrrr@{}}
\midrule
{Name} & \thead{Volume sent\\(US dollars)} & \thead{Volume received\\(US dollars)} & \thead{out\\degree}  &  \thead{in\\degree}  & \thead{Volume\\tot(US dollars)} \\

\hline
Abraxas Market   &  $29,822,178.9$ &  $23,044,463.2$ &       $21953$ &     $96612$ & $52,866,642.1$ \\
Acropolis Market    &      $11,196.7$ &     $11,407.6$ &  $101$ &        $201$ &    $22,604.3$ \\
Agora Market    &  $163,946,119.7$ & $148,224,155.3$ &      $122582$ &    $468708$ & $312,170,3$ \\
AlphaBay Market               & $605,445,951.5$ & $529,077,614$ &      $267818$ &    $1590672$ &  $1,134,523,565.2$ \\
Apollon Market  &       $17,384.5$ &      $15,113.6$ &  $57$ &        $138$ &  $32,498.1$ \\
Babylon Market    &     $144,292.6$ &     $149,257.5$ &   $902$ &       $1398$ &        $293,550.1$ \\
Berlusconi Market   &     $230,036.6$ &     $239,430.9$ &   $514$ &       $2153$ &        $469,467.5$ \\
Bilzerian24.net               &  $22,821,289.6$ &   $19,130,767.5$ &         $108$ &     $240232$ &     $41,952,057.1$\\
Black Bank Market     &  $14,841,938.8$ &  $13,858,325.9$ &       $15805$ &      $53260$ &     $28,700,264.8$ \\
Blue Sky Marketplace        &   $4,294,944.4$ &   $3,297,912.5$ &       $10210$ &      $16275$ &      $7,592,856.9$ \\
Dream Market    &  $78,031,896.0$ &  $60,049,434.3$ &       $46648$ &     $475260$ &    $138,081,330.3$ \\
East India Company Market     &   $3,638,096.5$ &   $2,942,049.9$ &  $4630$ &       $1951$ &       $6,580,146.4$ \\
Empire Market    &  $11,962,986.2$ &   $8,975,257.2$ &   $1309$ &      $66124$ &    $20,938,243.4$ \\
Evolution Market     &   $55,982,302.9$ &  $49,622,433.1$ &       $35415$ &     $219491$ &    $105,604,735.9$ \\
German Plaza Market  &   $1,032,802.5$ &     $951,757.3$ &          $22$ &      $10824$ &      $1,984,559.9$ \\
Hansa Market                  &  $62,087,671.5$ &  $61,171,541$ &       $73496$ &     $336045$ &    $123,259,212.5$ \\
House of Lions Market         &         $705.7$ &       $1,018.4$ &          $12$ &         $97$ &          $1,724.1$ \\
Hydra Marketplace             & $426,946,433.7$ & $474,549,308.6$ &      $113878$ &    $1081883$ &    $901,495,742.3$ \\
Middle Earth Marketplace      &   $9,861,173.8$ &   $8,549,901.3$ &        $9503$ &      $38506$ &     $18,411,075$ \\
Nucleus Market                &  $70,112,730.6$ &  $58,544,889.4$ &       $55522$ &     $207791$ &    $128,657,619.9$ \\
Olympus Market                &     $828,076.9$ &     $711,202.93$ &        $1877$ &       $4230$ &      $1,539,279.9$ \\
Oxygen Market                 &      $42,914.2$ &       $37,273.5$ &         $278$ &        $605$ &         $80,187.7$ \\
Pandora OpenMarket            &     $9,422,325.0$ &   $8,568,086.9$ &        $8864$ &      $35859$ & $17,990,411.9$ \\
Russian Anonymous Marketpl. & $131,000,457.9$ & $105,804,257.1$ &       $36794$ &     $745939$ &    $236,804,714.9$ \\
Sheep Marketplace             &  $15,624,992.4$ &  $11,624,434.9$ &        $7718$ &      $38612$ &     $27,249,427.4$ \\
Silk Road 2 Market            &  $85,610,718.5$ &  $70,325,928.9$ &       $48293$ &     $227239$ & $155,936,647.4$\\
Silk Road 3.1   &  $13,310,738.1$ &   $9,547,696.8$ &   $15574$ &      $64205$ &     $22,858,434.9$ \\
Silk Road Marketplace         & $172,812,766.4$ & $140,579,172.6$ &       $73114$ &     $400079$ &    $313,391,938.9$ \\
TradeRoute Market             &  $18,313,990.6$ &   $17,190,084.7$ &       $14318$ &     $104413$ &     $35,504,075.3$ \\
Unicc        & $147,418,817.2$ & $106,581,024.9$ &   $443$ &    $1301371$ &    $253,999,842.1$ \\
Wall Street Market            &  $68,596,630.4$ &  $52,623,050.2$ &       $26522$ &     $359656$ &     $121,219,680.6$ \\

\end{tabularx}
\caption{\label{tab:markets_sales} \textbf{Dark marketplaces overall activity.} The activity of the dark marketplaces as observed in our dataset. For each marketplace, the table reports the total volume sent and received by dark marketplace addresses. It also reports the total number of users who sent (in-degree) and received (out-degree) Bitcoins to/from dark marketplace addresses.}
\end{table}

\section{Moving Average Convergence Divergence Analysis}
\label{macd}
To further quantify the changes in dark marketplaces traded volume, we calculate the Moving Average Convergence Divergence (MACD) of the weekly trading volume. The MACD is a trading indicator used in stock marketplaces to quantify price movements and fluctuations. It is composed of three time series. Firstly, the MACD, calculated as the difference between the exponential weighted moving average of the trading volume for a period of $12$ weeks and the exponential weighted moving average of the trading volume for a period of $26$ weeks. Secondly, the signal line, computed as the $9$ weeks exponential weighted moving average of the MACD time series. Finally, the last time series, known as the histogram, representing the difference between the MACD and the signal line.

\begin{figure}[H]
\centering
\includegraphics[width=5in]{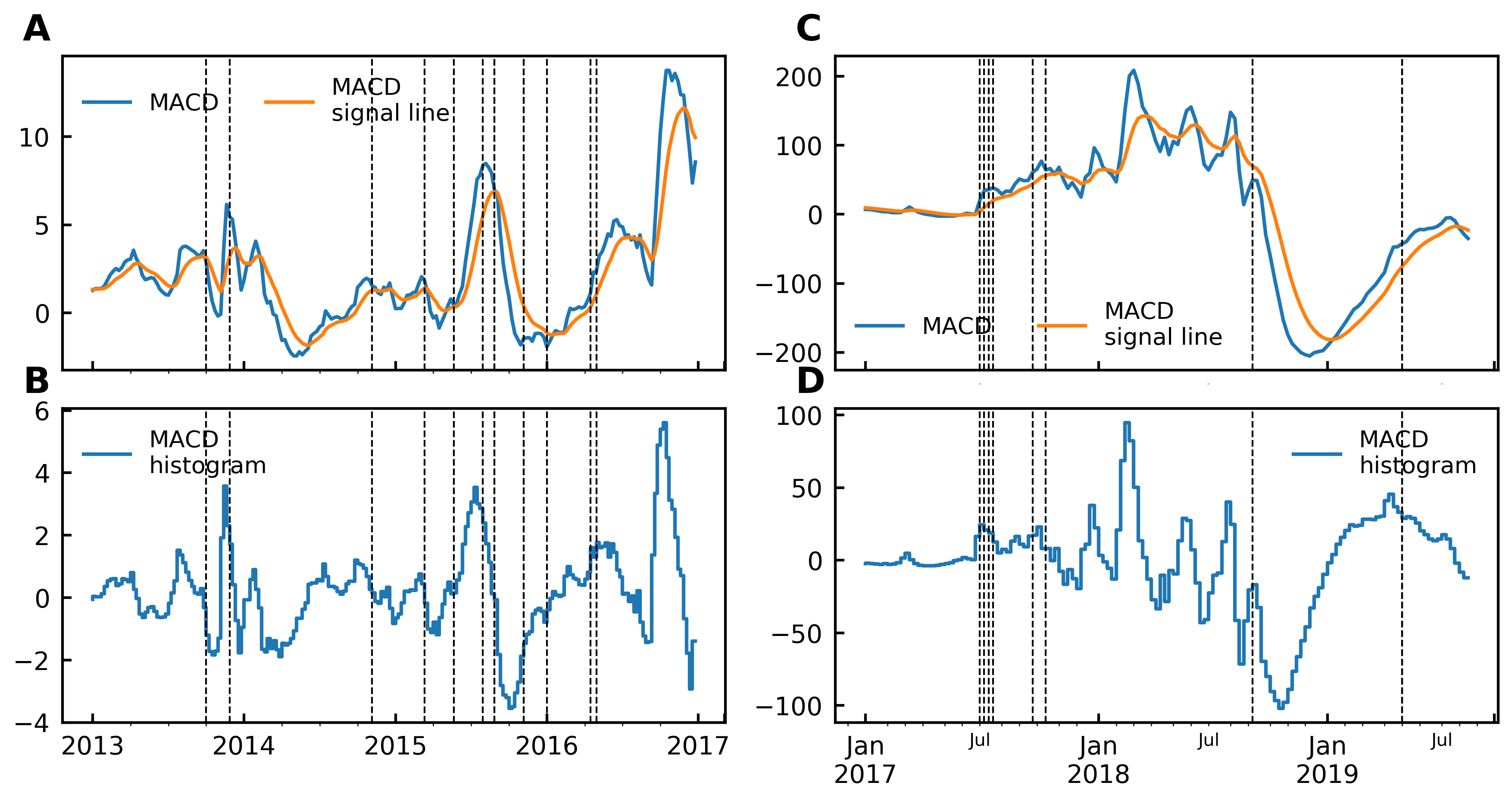} 
\caption{\textbf{Moving Average Convergence Divergence (MACD)} \textbf{(A)} The MACD (blue line) and MACD signal line (orange line) for dark marketplaces trading volume from $2013$ till $2016$. \textbf{(B)} The MACD histogram (blue line) for the dark marketplaces trading volume from $2013$ to the end of $2016$. \textbf{(C)} The MACD (blue line) and MACD (orange line) signal line for dark marketplaces trading volume from $2017$ until July, $2019$. \textbf{(D)} The MACD histogram (blue line) for the dark marketplaces trading volume from $2017$ until July, $2019$. Vertical dashed lines represent marketplaces closure.}
\label{fig:macd}
\end{figure}

Fig.~\ref{fig:macd} shows the indicator behaviour across time. For each closure, there is a fluctuation in the MACD line and the histogram line indicates a downward change in the overall dark marketplaces volume. However, an upward change can be observed after the closures indicating that dark marketplaces recover.

\section{Migrant and non migrants}
\label{appmig}

\begin{figure}[H]
\centering
\includegraphics[width=5in]{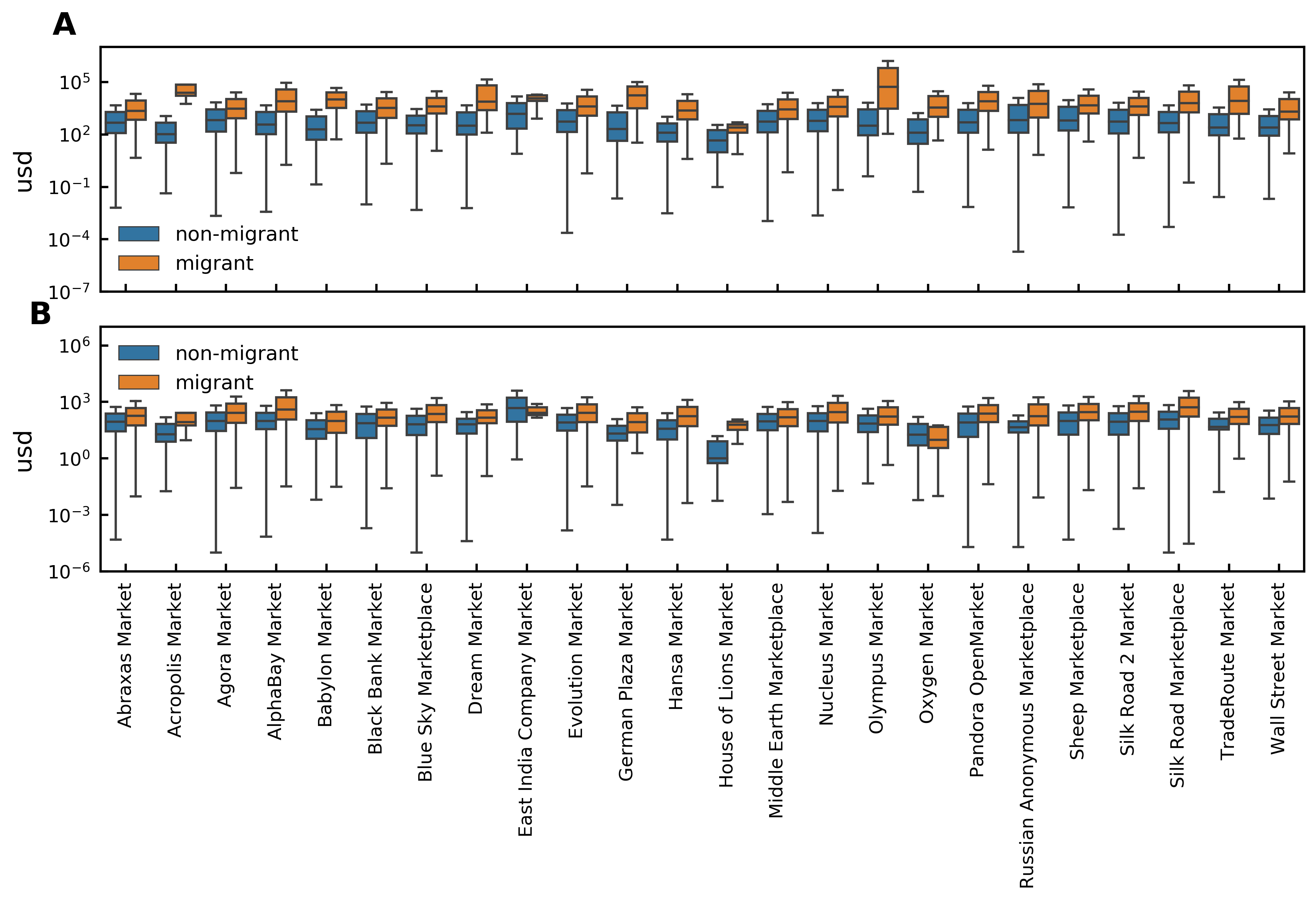}   
\caption{\textbf{Migrants are more active than other users.} \textbf{(A)} Total volume exchanged by migrant users (orange box-plots) and non-migrant users (blue box-plots) before the closure of their home marketplace. \textbf{(B)} Volume exchanged by migrant users (orange box-plots) and non-migrant users (blue box-plots) with their home marketplace. The horizontal line in each box represents the median. The lower box boundary shows the first quartile, and the upper one shows the third quartile. The whiskers show the minimum and maximum values within the $1.5$ lower and upper interquartile range.}
\label{fig:mig_beh_org}
\end{figure}

In the main text we show that for each closed marketplace, migrant users are more active in terms of the total amount they send and received overall, specifically with the closed dark marketplace. In this section, we show the behaviour across each closed marketplaces. Fig.~\ref{fig:mig_beh_org} shows that activity for migrants overall is higher than the non-migrants for each closed marketplace.

Table~\ref{tab_pvals} shows the results of a Kolmogorov smirnov test between the migrant and non migrant activity distribution.

\begin{table}
\renewcommand\arraystretch{1.5}
\centering
\begin{tabularx}{\columnwidth}{@{}Xrr@{}}
 \hline
\textbf{Dark marketplace} &  \textbf{$p$-value (dark marketplace transactions)}
 & \textbf{$p$-value (all transactions)}  \\
 \hline
 
Abraxas Market	& $5.9*10^{-85}$ &	$9.5673*10^243$  \\
Agora Market &	$0$ &	$0$   \\
AlphaBay Market	& $0$	& $0$   \\
Babylon Market	 & $3.794*10^{-04}$ &	$8.161*10^{-17}$   \\
Black Bank Market &	$1.632283*10^{-42}$ &	$1.735524*10^{-159}$  \\
Blue Sky Marketplace &	$1.138519*10^{-22}$ &	$6.731465*10^{-67}$  \\
Dream Market &	$7.749932*10^{-19}$ &	$1.204320e-66$  \\
Evolution Market &	$0$ &	$0$  \\
German Plaza Market &	$9.276236*10^{-18}$ &	$1.049758*10^{-44}$  \\
Hansa Market &	$4.727538*10^{-159}$ &	$0$  \\
Middle Earth Marketplace &	$9.356239*10^{-20}$ &	$2.203038*10^{-83}$  \\
Nucleus Market	& $2.538463*10^{-174}$ &	$6.319438*^{-268}$  \\
Olympus Market	& $1.453169*10^{-03}$ &	$1.647657*10^{-22}$  \\
Pandora OpenMarket &	$5.903384*10^{-65}$ &	$1.622666*10^{-187}$  \\
Russian Anonymous Marketplace &	$3.544511*10^{-83}$ &	$1.673279*10^{-48}$	\\
Sheep Marketplace	& $4.899846*10^{-112}$ &	$2.234014*10^{-231}$	\\
Silk Road 2 Market & 	$0$ &	$0$	\\
Silk Road Marketplace &	$0$ &	$0$ 	\\
TradeRoute Market	 & $1.563685*10^{-63}$ &	$3.078314*10^{-166}$	\\
Wall Street Market	& $8.111283*10^{-56}$ &	$1.109606*10^{-123}$ 	\\
\hline

\end{tabularx}
\caption{\label{tab_pvals}  \textbf{$P$ values between the migrants and stayers} The table shows the $p$ value results from the Kolmogorov smirnov test between the migrant and non migrants users distributions. The table report the results for the transactions to/from dark markets and the results for all the transactions.}
\end{table}

\end{document}